\documentclass{pasa}%

\usepackage{graphicx}
\usepackage{mathtools}
\usepackage{amsmath}
\usepackage[dvipsnames]{xcolor}
 
\definecolor{customColor}{RGB}{201, 124, 178}
\definecolor{customColor1}{RGB}{255, 178, 178}
\definecolor{customColor2}{RGB}{178, 178, 255}

\title{The Development of Non-coherent Passive Radar Techniques for Space Situational Awareness with the Murchison Widefield Array}

\author[Prabu et al.]{Prabu, S.$^{1,3}$, Hancock, P.$^1$, Zhang, X.$^2$ and Tingay, S.J.$^{1}$
\affil{$^1$International Centre for Radio Astronomy Research, Curtin University, Bentley, WA 6102, Australia}%
\affil{$^2$CSIRO Astronomy and Space Science, 26 Dick Perry Avenue, Kensington, WA 6151, Australia}
\affil{$^3$CSIRO Astronomy and Space Science, Corner Vimiera \& Pembroke Roads, Marsfield, NSW 2122, Australia}
}%

\jid{PASA}
\doi{10.1017/pas.\the\year.xxx}
\jyear{\the\year}

\usepackage{aas_macros}
\usepackage{hyperref} 
\hypersetup{colorlinks,citecolor=blue,linkcolor=blue,urlcolor=blue}

\begin{document}

\begin{frontmatter}
\maketitle

\begin{abstract}
The number of active and non active satellites in Earth orbit has dramatically increased in recent decades, requiring the development of novel surveillance techniques to monitor and track them.  In this paper, we build upon previous non-coherent passive radar space surveillance demonstrations undertaken using the Murchison Widefield Array (MWA).  We develop the concept of the Dynamic Signal to Noise Ratio Spectrum (DSNRS) in order to isolate signals of interest (reflections of FM transmissions of objects in orbit) and efficiently differentiate them from direct path reception events. We detect and track Alouette-2, ALOS, UKube-1, the International Space Station, and Duchifat-1 in this manner.  We also identified out-of-band transmissions from Duchifat-1 and UKube-1 using these techniques, demonstrating the MWA's capability to look for spurious transmissions from satellites. We identify an offset from the locations predicted by the cataloged orbital parameters for some of the satellites, demonstrating the potential of using MWA for satellite catalog maintenance. These results demonstrate the capability of the MWA for Space Situational Awareness and we describe future work in this area.
\end{abstract}

\end{frontmatter}

\section{INTRODUCTION }
\label{sec:intro}

The rising number of human-made objects in Earth orbit could lead to an increasing number of collision events between these objects. In the extreme limit, the Kessler effect \citep{1978JGR....83.2637K} is predicted to occur when the space density of objects increases beyond a critical value and a single collision leads to a cascade of collisions, rendering the entire orbit useless for future space missions \citep{Kessler2010KesslerPaper}. An average of 21 satellite collision warnings are issued by the US military every day \citep{2018Natur.561...24W} and most of these objects are in Low Earth Orbit (LEO) with speeds of approximately 8km/s. Many small satellites have very little on-board fuel for position-keeping and a single collision avoidance manoeuvre can reduce the operation life-time of the satellite drastically \citep{SCHAUB201566}. Also, many CubeSats and NanoSats do not have on-board propulsion systems and these satellites pose a threat to other operational satellites in orbit.

A well known example of these risks being realised was when two satellites (Iridium-33 and Kosmos-2251) collided in 2009  \citep{2009amos.confE...3K}. The density of objects in LEO has further increased in the past decade due to the testing of anti-satellite capabilities by China \citep{2007amos.confE..35K}, India \citep{akhmetov2019analysis}, Russia\citep{doi:10.1080/00963402.2019.1628458} and the United States \citep{PARDINI2009787}. These objects in orbit are monitored and cataloged by the Space Situational Awareness (SSA) \citep{Bobrinsky2010} program run by the European Space Agency (ESA) and the Space Surveillance Network (SSN) \citep{SSN} run by the US. Both these organisations have been expanding  their sensor networks in order to be able to detect and track multiple objects at any given time. However, due to companies such as SpaceX planning to launch mega-constellations \citep{2017AcAau.131...55R} in the future, a much larger number of sensors have to be utilised in order to accommodate the growth rate of satellites in LEO.

 Radio interferometers such as Long Wavelength Array (LWA) \citep{doi:10.1142/S2251171712500043,2014RaSc...49..157H} and LOw Frequency ARray (LOFAR) \citep{2013AandA...556A...2V,GAUSSIRANII20041375}  have detected reflections of non-cooperative transmitters from objects like meteors and aircrafts using correlated data in the past. In this work, we explore the use of the Murchison Widefield Array (MWA) for passive space surveillance in the FM band, building upon previous work. The transmitter-target-receiver architecture used in this work is similar to the GRAVES radar\footnote{\url{https://www.onera.fr/en/news/graves-space-surveillance-system}} except that we use interferometric correlated data and non-cooperative terrestrial FM transmitters.

The MWA is a low frequency ($70 - 300$\, MHz) radio interferometer built as a precursor to the Square Kilometre Array (SKA)  \citep{Tingay2013TheFrequencies}, located at the radio-quiet Murchison Radio-astronomy Observatory (MRO) in Western Australia. The MRO is one of two sites where the SKA will be built in the future. While the MWA has been designed and built with a primary mission for astrophysics and cosmology \citep{2013PASA...30...31B, 2019arXiv191002895B}, the facility has been shown to be a novel and effective instrument for SSA studies, utilised as part of passive radar systems that use terrestrial FM broadcasts as non-cooperative illuminators of opportunity for objects in Earth orbit.  The MWA represents a highly sensitive receiver of FM signals reflected off orbiting objects.

The MWA has been used previously to detect the International Space Station (ISS) by using non-coherent \citep{Tingay2013OnFeasibility} and coherent \citep{7944483} passive radar detection techniques. In coherent detection, the signal transmitted by the FM station is used as a reference for designing matched filters to search for reflection events, while in non-coherent detection, the search is done using interferometer correlated data in the image domain.  In this paper, we focus on the development of non-coherent  techniques for space surveillance by the MWA.  The development of coherent techniques is described by \citep{8835821} and in an upcoming article by Hennessy et al. (2019, IEEE submitted).

\citet{Tingay2013OnFeasibility} carried out observations during the commissioning phase of the MWA as a proof-of-concept for the non-coherent passive radar technique. This work  detected the ISS, using the transmissions reflected off the ISS from a variety of terrestrial FM radio stations. \citet{Tingay2013OnFeasibility} also published electromagnetic simulations, predicting that debris with a radius greater than $0.5$\,m could be detected by the MWA via FM reflections at ranges up to $1000$\,km, with a $1$\,s integration time and a $50$\,kHz bandwidth.

Using similar techniques, the MWA has also been used to search for meteors \citep{Zhang2018LimitsMWA}, based on possible intrinsic radio emission, as seen at lower frequencies by the Long Wavelength Array \citep{2012JAI.....150004T,2013ITAP...61.2540E,2016JGRA..121.6808O}, or the reflection of FM radio waves from the ionisation trails left by meteors.  In \cite{Zhang2018LimitsMWA}, the static celestial sources in their images were removed using difference imaging techniques, leaving non-static meteors visible as transient signals in the difference images. During these observations, some apparent FM reflections from satellites were also detected (of objects much smaller than the ISS) but were not investigated in detail.  The current paper undertakes a comprehensive examination of these detections and extends the difference imaging techniques of \citet{Zhang2018LimitsMWA}, to improve the detection of objects in Earth orbit for the purposes of SSA using non-coherent passive radar techniques with the MWA.

This paper is compiled as follows. In Section 2 we discuss the observations and the data processing analysis. Section 3 describes the results obtained from our analysis. The results and conclusions are discussed in Sections 4 and 5.

\section{Observations and Data Processing}

\subsection{Observations}
All of the observations we have utilized are zenith pointing drift scans from Phase 1 of the MWA (128 tiles distributed over a $\sim3$\,km diameter area), observing the sky at $72.335-103.015$\,MHz (arranged as 24 $\times$ 1.28 MHz coarse channels). Table \ref{tab1} contains the list of the dates and times of these target observations and the identification of the objects detected (along with some characteristics of those objects).  Also listed in Table \ref{tab1} are the calibrator sources associated with the observations. 

\begin{table*}[h!]
\caption{List of observations and identified target objects within those observations.}
\centering
\begin{tabular*}{\textwidth}{@{} c\x c\x c\x c\x c\x c\x c\x c\x c\x c\x@{}}
\hline \hline
 Observation ID    &Date      &  Start        &  Stop 
          & Target/ &   Range to Target     & RCS$^{a}$ & Required TLE Offset \\
~     &UT        & UT            & UT 
          & Calibrator&          (km)            & (m$^{2}$) & (s)\\ 
\hline \hline
 1102604896&2014-12-14 & 15:07:58 & 15:11:50 &  Alouette-2 & 2184.9 - 2298.0 & 0.985  & 9.0\\
  1102627456&2014-12-14   &21:23:58  &21:25:50  &Hydra A&    & \\ \hline
 1142340880 &2016-03-18 & 12:54:22 & 12:58:14 & ALOS &712.8 - 751.8 &  13.593 & 1.0\\ 
  1142332016& 2016-03-18 &10:26:38 & 10:29:34&Pictor A& &    \\ \hline
 1142351440 &2016-03-18 & 15:50:22 & 15:54:14 & UKube-1 &  653.3 - 688.8  &    0.118 & 1.75\\   
   1142332016& 2016-03-18 &10:26:38 & 10:29:34&Pictor A&   &\\ \hline
 1142425368&2016-03-19 & 12:22:30 & 12:26:22 & ISS & 437.4 - 501.2  & 399.052& 0.0 \\ 
  1142418344& 2016-03-19  & 10:25:26 &10:28:22 &Pictor A &  & \\ \hline
 1142521608&2016-03-20 & 15:06:30 & 15:10:22 & Duchifat-1& 623.3 - 710.8 &  0.037 & 1.25\\ 
  1142504680 & 2016-03-20 &  10:24:22 &  10:27:18 &Pictor A  &  & \\ \hline
\hline \hline
\end{tabular*}\label{tab1}
\tabnote{$^{a}$Radar Cross Section (RCS) obtained From \url{https://celestrak.com/pub/satcat.txt} \\
All the above targets/calibrator were observed at $72.335-103.015$\,MHz. The range to target is the maximum and minimum line of sight distance at which the target was detected. Alouette-2 is also seen briefly in the observation 1102605136 (spanning from UTC 2014-12-14 15:12:00.00 to 2014-12-14 15:15:52.00), but was not used for the analysis performed in this paper.}
\end{table*}

\subsection{Data Processing}
The visibility data for the observations in Table \ref{tab1} were downloaded as measurement sets \citep{2007ASPC..376..127M} from the MWA node of the All-Sky Virtual Observatory\footnote{\url{https://asvo.mwatelescope.org/dashboard}} (ASVO). Time averaging of $2$\,s ($4$\,s for observations containing the Alouette-2 satellite, for reasons explained later) was used along with frequency averaging of $40$\,kHz. The first and last $80$\,kHz, along with the central $40$\,kHz of every 1.28 MHz coarse frequency channel was flagged due to the characteristics of the band-pass filter. The ASVO uses COTTER \citep{2015PASA...32....8O} to convert native MWA format visibility files to measurement sets. RFI detection in COTTER was disabled when retrieving the target observations, so that the signals of interest were not automatically flagged. However, ASVO does apply the hardware flagging to data that is performed on-site. 

The target observations were calibrated using calibration observations from the same night. The calibration observations were retrieved from the ASVO using the same parameters as the target observations but with RFI detection enabled in COTTER, in order to obtain good calibration solutions. The calibration observations were additionally pre-processed using AOFLAGGER \citep{2015PASA...32....8O} to flag all the baselines in time and frequency with RFI in them.

\begin{figure}[h]
\begin{center}
\includegraphics[width=\linewidth,keepaspectratio]{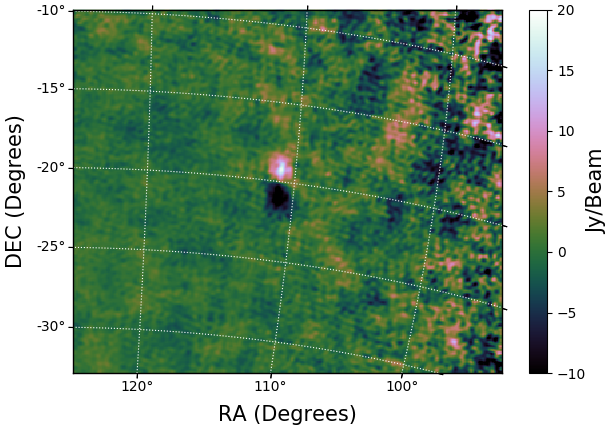}
\caption{Primary beam corrected $30.72$\,MHz bandwidth difference image of ALOS centered at $87.675$\,MHz. ALOS is a remote sensing satellite orbiting at an altitude of about $690$\,km and has an RCS of $13.6$\,$m^2$. The satellite also has large solar panels, that when fully deployed have an RCS of $66.0$\,$m^2$.}
\label{Fig1ALOS}
\end{center}
\end{figure}

\begin{figure}[h]
\begin{center}
\includegraphics[width=\linewidth, keepaspectratio]{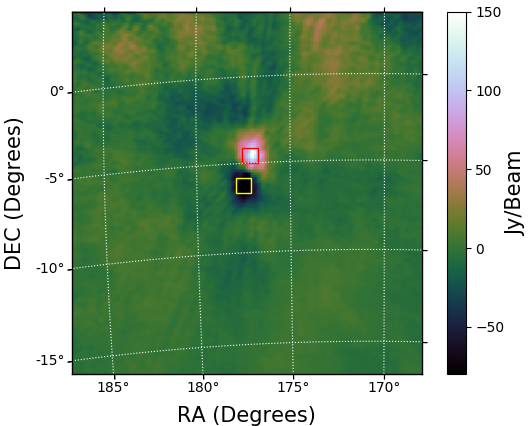}
\caption{Primary beam corrected $30.72$\,MHz bandwidth difference image of UKube-1 centered at $87.675$\,MHz. UKube-1 is a 3 Unit CubeSat. The figure also shows the box make by the automated DSNRS script used for integrating flux density in the head and the tail of the streak.}
\label{Fig2UKUBE}
% Use labels that are more descriptive and dynamic: eg \label{fig:diff_ukube}
\end{center}
\end{figure}

%%%%%%%%%%%%%%%%%%%%%%%%%%%%%%%%%%%%%%%%%%%%%%%%%%%

When generating calibration solutions from the calibration observations, solutions for the times and frequencies flagged due to RFI were determined by interpolating the solutions across these times and frequencies. This was essential as we did not want the flags due to RFI in the calibration observations to be carried across to the target observations during calibration transfer. 

Once the target observations were calibrated, they were imaged using WSCLEAN \citep{offringa-wsclean-2014,offringa-wsclean-2017} at every $2$\,s timestep ($4$\,s for Alouette-2) and at every $40$\,kHz fine frequency channel. 
In interferometer theory, a source is considered to be in the far-field if the received wave-front is planar as seen by a baseline of length $D$. 
The transition between the near-field and the far-field (the Fraunhofer distance) is given by $d=2D^{2}/\lambda$ where $\lambda$ is the observing wavelength.
Some satellites considered in this work are within a few hundred kilometers of the MWA, which puts them into the near field as seen by the MWA's longest baselines ($3$\,km). 
Consequently we restrict our analysis to baselines shorter than $500$\,m in order to avoid near-field affects \citep{Zhang2018LimitsMWA}.
Natural weighting was also used for all objects. CLEAN was not used, as the step after imaging is to form difference images, which removes the celestial sources and their side-lobes. Pseudo Stokes I images (i.e, without the primary beam correction) were made from the data and were used for the analysis described in \S2.3.

We also produced images of the full $30.72$\,MHz bandwidth using multi-frequency synthesis, again at every time step. However, the full bandwidth images combine lots of channels with no signal and reduce the signal to noise ratio. Hence, we use these images only for preliminary detection (and position verification) as manual inspection of difference images from every fine channel was not feasible.  These images were made in XX and YY polarisations, which were primary beam corrected before combination, to produce a primary beam corrected Stokes I, full bandwidth image.

The data reduction pipeline used in this work incorporates the difference imaging technique that was found to work effectively by  \citet{Zhang2018LimitsMWA}. Once imaged, difference images were formed at each time step $t$ by subtracting the image at time step $t-1$ from the image at time step $t$. The difference images remove the persistent celestial sources, along with their side-lobes, thus greatly reducing the side-lobe confusion noise in the difference images. The difference images reveal objects that move rapidly in Right Ascension and Declination (such as satellites, orbiting debris, planes, and long duration meteors) as streaks, characterised by a positive intensity head (in the direction of motion of the object) and a negative intensity tail. The phase centres of these images were fixed at the pointing centres of the MWA beam for the observations (zenith in this case).  Examples of difference images revealing such streaks due to the objects listed in Table \ref{tab1} are shown in Figures \ref{Fig1ALOS} and \ref{Fig2UKUBE}.

\begin{figure}[h!]
\begin{center}
\includegraphics[width=\columnwidth, keepaspectratio]{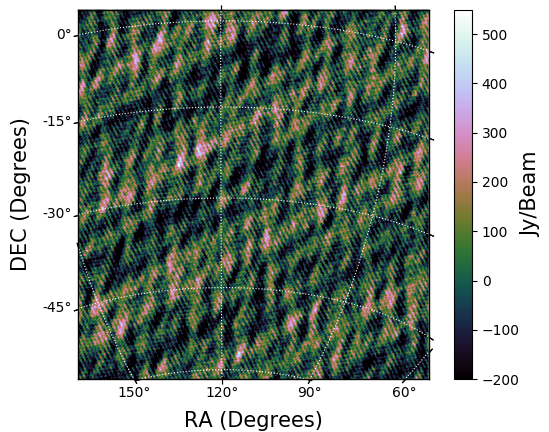}
\caption{Difference image for one $40$\,kHz frequency channel with direct FM reception.}\label{Fig3DiffImage}
\end{center}
\end{figure}

\begin{figure}[h!]
\begin{center}
\includegraphics[width=\columnwidth,keepaspectratio]{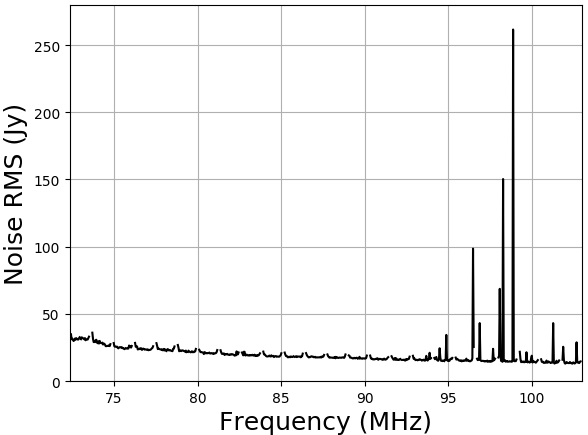}
\caption{Plot showing the variation of noise RMS of difference images with frequency. Note that the plot is discontinuous at the center and edge of every coarse channel due to flagging. }
\label{Fig4NOISE}
\end{center}
\end{figure}

\subsection{Dynamic Signal to Noise Ratio Spectrum (DSNRS) Analysis}

During the development of the imaging and difference imaging methodology, we noticed that the images could be affected by FM signals in two ways.  First, the signal of interest was present, that being the FM signals reflected off objects in orbit and thus confined to a small region of the image plane.  Second, FM signals could enter the MWA field-of-view by virtue of direct reception from the transmitter, via atmospheric ducting.

When FM reflections are present in an image, the signals are highly localised in the image, corresponding to the locations of the objects reflecting the signals.  In this case, the overall RMS in the image is very close to that of a thermal noise dominated image with no signal present.  

However, when an image is affected by direct reception of FM signals, the overall image RMS is greatly increased relative to a thermal noise dominated image.  For example, we show a difference image for one $40$\,kHz frequency channel affected by direct reception, and the variation of image RMS in difference images as a function of frequency, for one of the observations in Figures \ref{Fig3DiffImage} and \ref{Fig4NOISE}, respectively.

We utilised these characteristics to distinguish between reflected and direct reception FM signals in our data and to isolate the signals of interest, as follows. 

The archived Two Line Element (TLE) data\footnote{TLE data obtained from \url{https://space-track.org}} were obtained for the epochs at which the observations were made, for the relevant objects. The TLE data contain the orbital parameters at a given epoch along with the satellite ID. The Ephem\footnote{\url{https://pypi.org/project/ephem/3.7.3/}} python module was used to propagate the satellite using TLE data to the UTC time of the difference images. 

%The Ephem module enables a user to create an observer class object by providing the latitude, longitude, and elevation of the MWA, returning the azimuth and elevation of the satellite at a given epoch as seen by the observer (the MWA in this case). The World Coordinate System\footnote{\url{http://docs.astropy.org/en/stable/wcs/#module-astropy.wcs}} (WCS)  from the difference image header was used to convert azimuth and elevation of the satellite as seen by the MWA to Right Ascension and Declination and (X,Y) pixel location for that difference image.

\begin{figure*}[h!]
\begin{center}
\includegraphics[width=\linewidth,height=25pc]{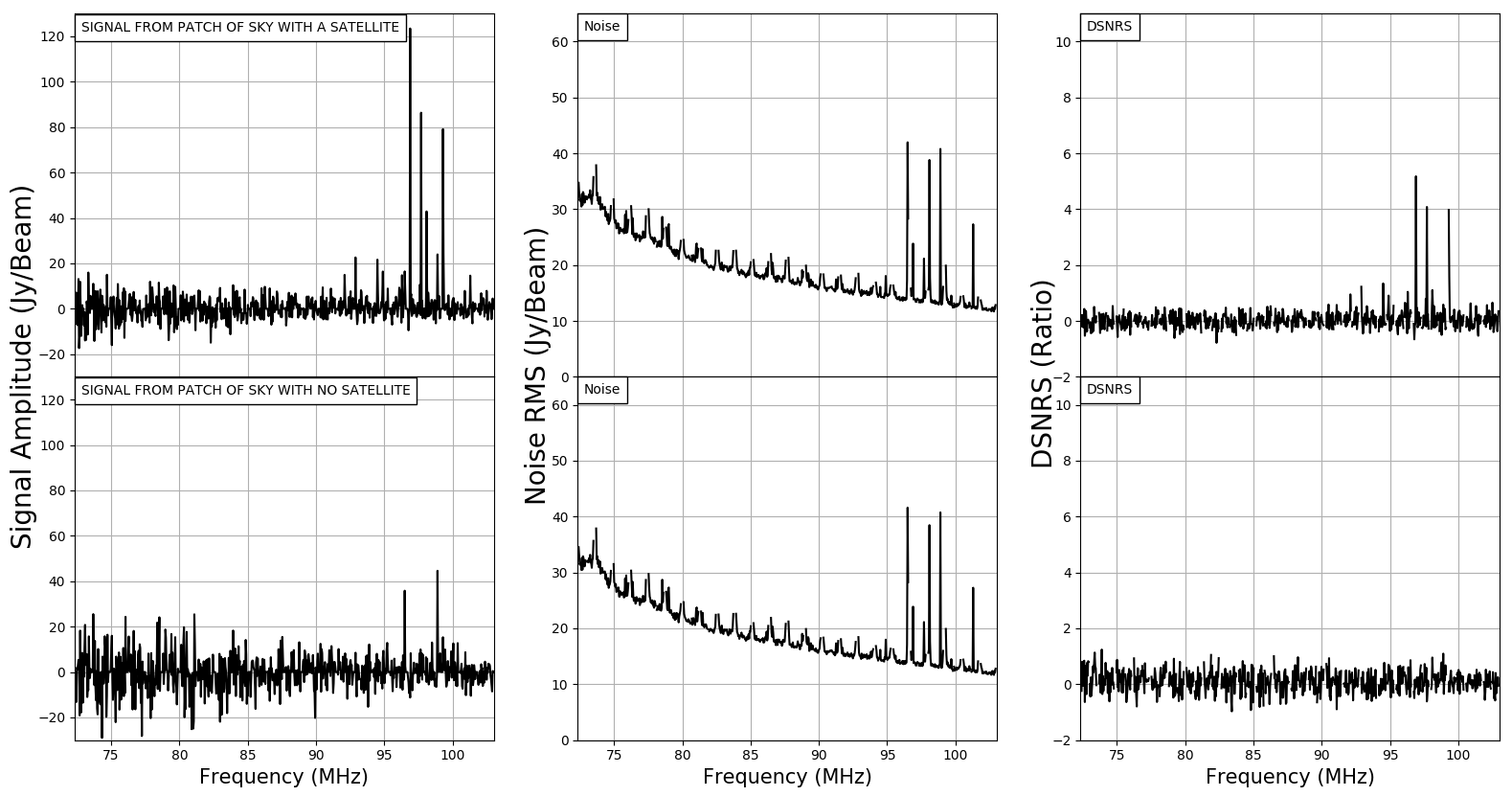}
\caption{The left, middle and right panel show the numerator, denominator and the resultant value of Equation \ref{E1}  when applied on a part of the sky with (top) and without (bottom) a satellite. Note the plot is discontinuous as the center and edge of every course channel due to flagging. }
\label{Fig5DSNRSUTILITY}
\end{center}
\end{figure*}

The predicted satellite location was used to form boxes around the head and the tail of the streak in the difference images, as shown in Figure \ref{Fig2UKUBE}. However, the predicted satellite position did not always exactly match with the MWA detections and time delays of a few seconds (given in Table \ref{tab1})  had to be provided to the Ephem module in order to make the predicted position coincide with the detection (note that the offset is not due to error in the instrument but due to the TLE for the satellites being outdated during the observation). The sizes of the boxes were calculated using the distance the satellite was predicted to move in the image plane, and to contain the majority of the signal at all times. These boxes were used to calculate the Dynamic Signal to Noise Ratio Spectrum (DSNRS) for all of the satellites. Due to the motion of these satellites being resolved, and the signal of interest being contained in the positive "head" and the negative "tail", the DSNRS describes the mean signal contained in the boxes, divided by the RMS of the noise calculated in the image away from the signal, and is presented in Equation \ref{E1}.

%Due to the satellites not being detected at all time-steps and channels, classification of the satellites as an "island" (instead of box) was not possible, as the "island" is not defined in the difference images with no  detection and the island is also different between frequencies it is detected. Hence, we use a box kernel that contains the majority of the signal at all times to plot the mean SNR of the streak with time and frequency, at the cost of not  containing the entire signal included inside the box. DSNRS is derived as follows. 

\begin{equation}
    \begin{multlined}
         DSNRS(t,f) = \frac{\frac{\sum_{i=1}^N J_{Head}(t,f)-\sum_{k=1}^MJ_{tail}(t,f)}{M+N}}{RMS(t,f)}\\\\
         = \frac{\sum_{i=1}^N J_{Head}(t,f)-\sum_{k=1}^M J_{tail}(t,f)}{RMS(t,f)\times(M+N)}
    \end{multlined}
    \label{E1}
\end{equation}

%\begin{equation}
%    \begin{multlined}
%         DSNRS(t,f) = \frac{Mean Signal From %Streak}{Noise}\\ \\
%         = \frac{\frac{\sum_{i=1}^N %J_{Head}(t,f)-\sum_{k=1}^MJ_{tail}(t,f)}{M+N}}{RMS(t,f)%}\\\\
%         = \frac{\sum_{i=1}^N %J_{Head}(t,f)-\sum_{k=1}^M %J_{tail}(t,f)}{RMS(t,f)\times(M+N)}
%    \end{multlined}
%    \label{E1}
%\end{equation}

%\begin{align}
%    DSNRS(t,f) = \frac{Mean Signal From Streak}{Noise}\\
%   = lkjdsf
%
%    DSNRS(t,f) =\frac{\sum_{i=1}^N J_{Head}(t,f) - %\sum_{k=1}^M J_{tail}(t,f)}{RMS(t,f)\times (M+ N)}
%    \label{E1}
%\end{align}

where $J_{Head}$ and $J_{Tail}$ are the intensity (per pixel values) in the tail and the head of the streak in a difference image, respectively; $N$ and $M$ are the number of pixels in the head and the tail, respectively. RMS is the root mean square of the difference image calculated at time step $t$ and at frequency $f$, at a region in the image containing no satellites. We negate the tail summation term in the above equation as the signal in the tail is negative. Both the numerator and denominator in Equation \ref{E1} have the dimensions of intensity (Jansky/beam), thus the resultant value of Equation \ref{E1} is a dimensionless number that varies with time and frequency, hence the use of dynamic spectrum in our terminology. Note that we utilise the DSNRS as a qualitative detection metric, to identify the frequencies reflected by the satellite and to isolate those signals in image, time, and frequency space.  The DSNRS metric, while a measure of signal-to-noise, does not imply any particular adherence to an underlying statistic.  The noise term contains both Gaussian and complicated non-Gaussian components.

%and is analogous to a signal to noise ratio.  Thus, we term Equation \ref{E1} as the Dynamic Signal to Noise Ratio Spectrum (DSNRS).

Figure \ref{Fig5DSNRSUTILITY} shows the utility of the DSNRS in isolating the reflected FM signals of interest in our difference images.  In the bottom three panels of Figure \ref{Fig5DSNRSUTILITY}, DSNRS was applied to a randomly selected location on the sky that did not contain a satellite.  The bottom-left panel of Figure \ref{Fig5DSNRSUTILITY} shows the summed intensities in head and tail boxes.  The bottom-middle panel of Figure \ref{Fig5DSNRSUTILITY} shows the image RMS.  The bottom-right panel of Figure \ref{Fig5DSNRSUTILITY} shows the result of DSNRS, that all of the signal found at the randomly selected location is due to direct reception of FM signals by the MWA.

However, the top three panels of Figure \ref{Fig5DSNRSUTILITY} shows the same observation, but with the top-left panel showing the summed intensities in head and tail boxes selected to correspond to a known satellite.  The top-right panel of Figure \ref{Fig5DSNRSUTILITY} shows the result of DSNRS, that the FM signals reflected from the satellite are isolated.  Thus, using DSNRS, we can determine the time and frequency dependence of the reflected FM signals, distinguished from direct reception FM signals.

\begin{figure*}[h!]
\begin{center}
\includegraphics[width=\linewidth, height=80pc, keepaspectratio]{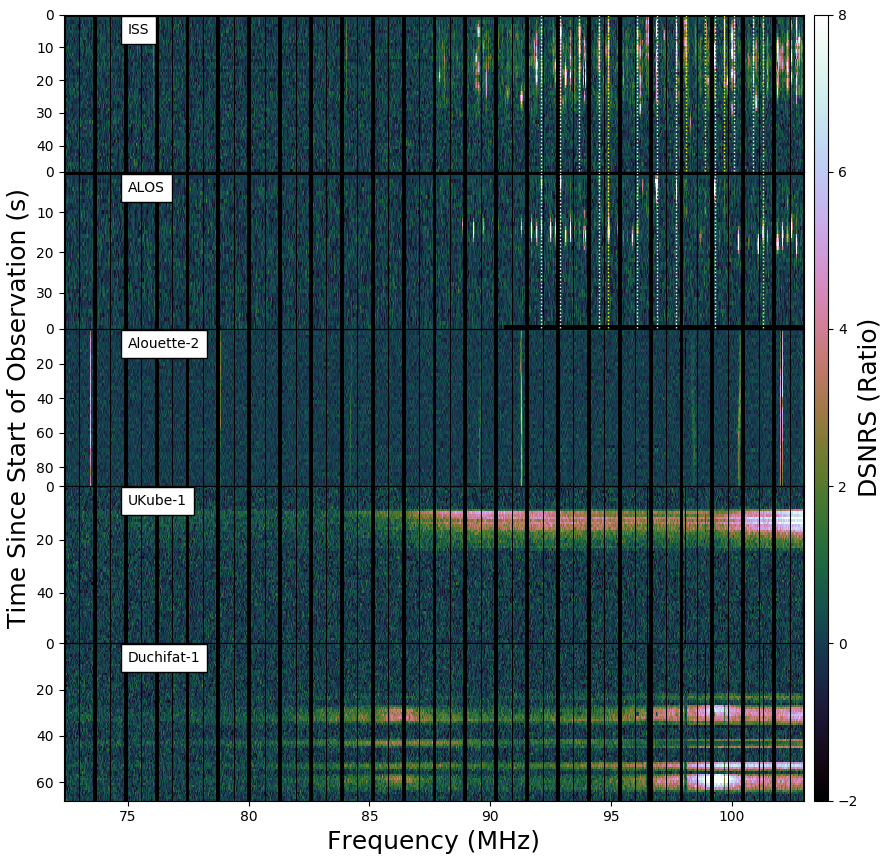}
\caption{DSNRS plots of all the targeted objects mentioned in Table \ref{tab1}. The edge and middle of every course channel was flagged (represented by black lines) while the other vertical and horizontal flags are due to missing visibilities caused by hardware failure. The top two panels have dotted white and yellow lines showing the fine channels reflecting FM transmitters from Perth and Geraldton, respectively. Note that the maximum values of the DSNRS plots for the CubeSats are much greater than 8 but the colobar has been clipped between -2 and 8 in order to accommodate reflecting and transmitting satellites in the same figure.}
\label{Fig6}
\end{center}
\end{figure*}

The method also proved effective in revealing the reflections, even in frequency channels that contained satellite reflections as well as direct FM reception in them. This was tested by superimposing a difference image corrupted by direct FM reception (such as the one shown in Figure \ref{Fig3DiffImage}) over the difference image obtained in every other frequency channel. The FM reflection of the RFI was revealed even in the presence of increased noise, but the amplitude was attenuated. The DSNRS method was automated for the detection of reflections from all the frequency channels and time-steps for a given satellite by creating boxes at the predicted locations of the satellite, as shown in Figure \ref{Fig2UKUBE}. The results of the DSNRS analysis are given in \S3.1. 

\section{Results}

\subsection{Results of DSNRS Analysis for Targeted Objects}
The DSNRSs for the objects listed in Table \ref{tab1} are shown in Figure \ref{Fig6}. From these figures, it can be seen that many transmitted signals from different locations are reflected by ALOS and the ISS. The ISS also has a reflection at $87.8$\,MHz which must be from a FM transmitter outside of Australia, given it is outside the allocated frequency range for FM broadcasts in Australia. It can be also noted that the ISS and ALOS have many common frequencies, presumably because they are quite similar in ranges (thus having similar reflection geometries between transmitter, object, and receiver). The brightening and fading of the signal is likely due to the changing transmitter-object-receiver geometry and/or Radar Cross Section (RCS) as the object moves across the sky. 

On the other hand, Alouette-2, whose altitude varies from $500-2700$\,km, was detected at a maximum line of sight distance of $2298$\,km and reflects signals from transmitters very distant (Loxton, Mildura, and Melbourne) from the MWA. The reflected signals for Alouette-2 are quite stable with time (unlike that of ISS and ALOS), perhaps due to it having a much slower angular speed across the sky due to its higher altitude (this is why $4$\,s time steps were used for Alouette-2, rather than $2$\,s time steps). Also the RCS of Alouette-2 may not change much relative to the transmitter/receiver geometry, due to it being almost spherical in shape (a perfect test particle for radar studies!). Alouette-2 appears to have a frequency reflected at approximately $73.4$\,MHz, again outside the FM broadcast band in Australia.  This frequency overlaps with the VHF-low analog TV broadcast band, but these broadcasts have a much broader bandwidth than seen in the Alouette-2 DSNRS.  Thus, the transmitter responsible for this reflection is unidentified.

\begin{figure}[h!]
\begin{center}
\includegraphics[width=\columnwidth]{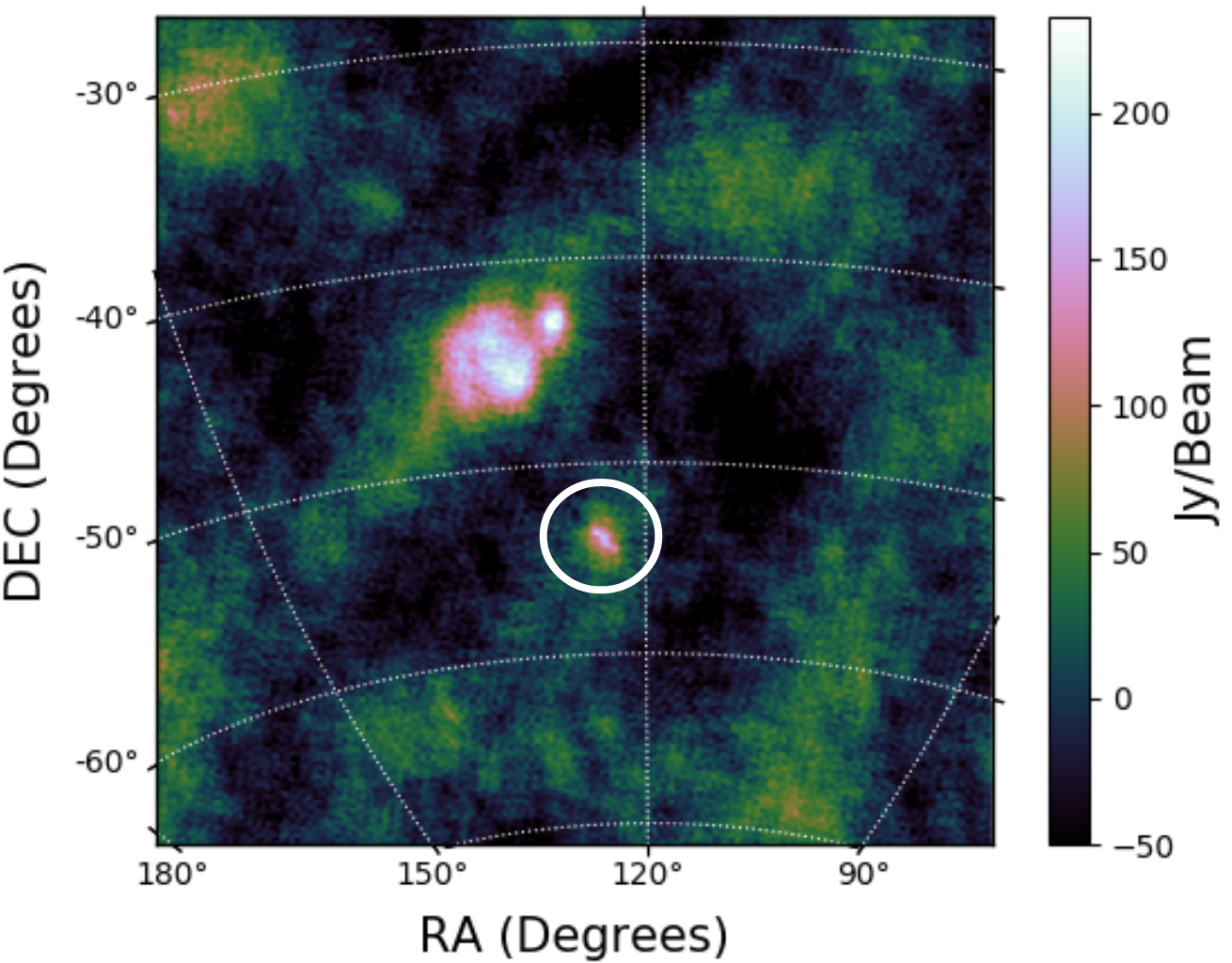}
\caption{The bright spot inside the white circle is the ISS as seen in a single $40$\,kHz fine channel dirty image. The diffuse structure in the image is the Vela supernova remnant.}
\label{Fig7ISSFINECHANNEL}
\end{center}
\end{figure}

\begin{figure}[h!]
\begin{center}
\includegraphics[width=\columnwidth]{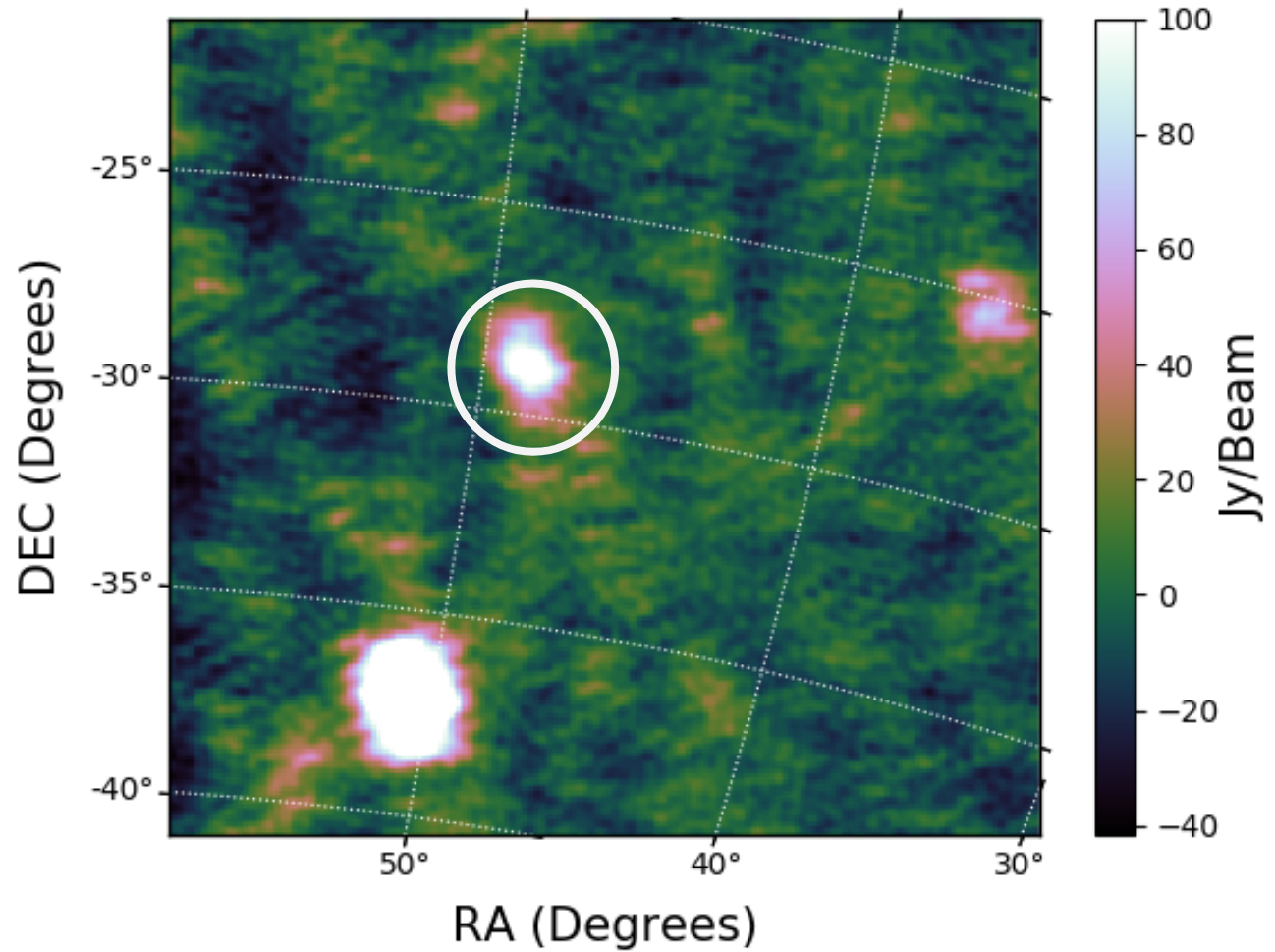}
\caption{Alouette-2 as seen in a single $40$\,kHz fine channel dirty image. The source in the bottom left is Fornax-A and the bright spot in the right is a cluster of different sources seen as a single emission region due to using baselines shorter than $500$\,m.}
\label{Fig8ALOUETTEFINECHANNEL}
\end{center}
\end{figure}

When the frequency channel for imaging was selected appropriately, these satellites appeared in standard images even without making difference images or performing CLEAN. The fine channels that appeared the brightest in the DSNRS plots in Figure \ref{Fig6} were chosen to make these images. The ISS and satellites as distant as Alouette-2 appeared well above the side-lobe confusion noise in these channels (about $12$\,Jy), as shown in Figure \ref{Fig7ISSFINECHANNEL} and Figure \ref{Fig8ALOUETTEFINECHANNEL}, respectively.

BANE (part of AegeanTools) \citep{Hancock_Aegean_2012,Hancock_Aegean2_2018} was used to perform background noise estimation and the calculation of flux densities (integrated intensities) from these images. The Effective Isotropic Radiated Power (EIRP) power calculated for Alouette-2 was found to be about $11$\,$\mu$W(at a range of $2241$\,km), which is significantly higher than the power predicted (about $20$\,pW) for a $1$\,m diameter sphere using an XFdtd simulation in \cite{Tingay2013OnFeasibility}.  This could be due to the presence of antennas on the satellite making it a good reflector (due to increased RCS) or due to the satellite reflecting FM stations other than those considered in the simulation.

Two of the detected satellites are CubeSats (Duchifat-1 and UKube-1) and they appeared far brighter than large satellites such as the ISS and ALOS. Via the DSNRS analysis, these CubeSats were found to be most likely transmitting in the FM band, rather than reflecting terrestrial broadcast signals.  This is evidenced in Figure \ref{Fig6} by the broadband nature of the signals from the CubeSats and the lack of identifiable narrow band FM signals associated with FM reflections. Amateur satellites are allowed to transmit between $144-146$\,MHz for down-link telemetry purposes but it appears that these CubeSats are producing significant transmitted power outside this allocated telemetry band. The measured EIRP from these CubeSats is approximately $256$\,mW.   Note that EIRP is calculated assuming the transmission is isotropic. If the CubeSat transmission is directional in nature, then the actual transmitted power would be lower than indicated by the EIRP calculation. Also, since the CubeSats are smaller than the wavelength considered here, the out-of-band transmission could be due to failed or defective hardware.

\begin{figure}[h!]
\begin{center}
\includegraphics[width=\linewidth, keepaspectratio]{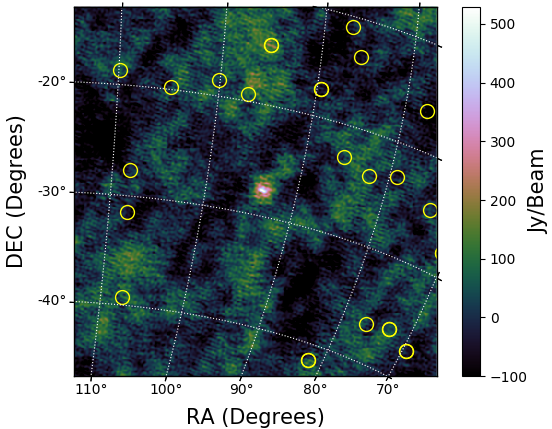}
\caption{An object not in the TLE catalog. The yellow circles are the location of cataloged orbiting objects at that epoch. Note that the object does not appear as a streak due to the signal being confined within the $4$\,s used in the difference image.}
\label{Fig9UNCATALOG}
\end{center}
\end{figure}

\begin{figure}[h!]
\begin{center}
\includegraphics[width=\columnwidth]{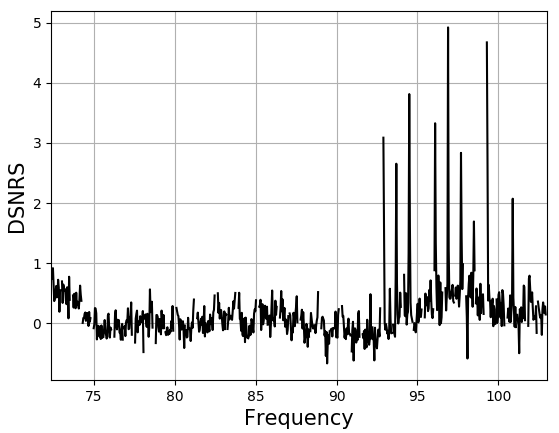}
\caption{A single time-step DSNRS plotted for the short duration signal seen in Figure \ref{Fig9UNCATALOG}.}
\label{Fig1ALOS0}
\end{center}
\end{figure}

\subsection{Detection of Additional Signals}
In the process of examining the difference images for the targeted objects, at the high time and frequency resolution required to calculate the DSNRS, we noted that our data contained additional, generally short duration, signals.  A significant number of these signals were detected, with a range of characteristics.  The examination of the details of these short duration signals is beyond the scope of this current publication but we show an example here and will present a detailed study of these additional signals in a future publication.

Figure \ref{Fig9UNCATALOG} shows a difference image in which a signal appears for only one difference time-step ($4$\,s). The lack of a negative tail confines the event to being shorter than $4$\,s. The detection is significant and the DSNRS shows that it is due to reflected FM signals (Figure \ref{Fig1ALOS0}).  However, Figure \ref{Fig9UNCATALOG} also shows that the signal does not correspond with the position of any catalogued satellite.

Most likely this signal is due to a reflection from a meteor trail, but other reasons for a short duration signal may be a highly variable RCS, due to a rapidly tumbling object or a geometrically complicated object.

\section{DISCUSSION}
\subsection{Completeness}
In this work, we perform a basic demonstration of the techniques involved for performing space surveillance with the MWA and in future we will develop it into a more sensitive blind detection pipeline. Also, due to having just 5 positive detections in the observations used in this work, we only perform a basic completeness check as mentioned below.

\begin{figure}[h!]
\begin{center}
\includegraphics[width=\columnwidth]{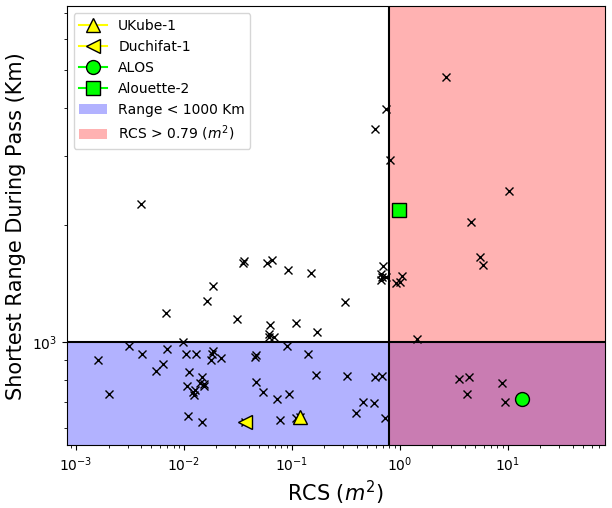}
\caption{Satellites/debris that passed through the half power beam during the observations mentioned in Table \ref{tab1}. The transmitting satellites are shown in yellow and the reflecting satellites are shown in green. The region shown in  \fcolorbox{black}{customColor}{\rule{0pt}{1pt}\rule{1pt}{0pt}} is the detection parameter space for MWA in FM frequencies. Note that ISS is not part of the above figure, due to it being detected outside the half power beam.}
\label{Fig11}
\end{center}
\end{figure}

 \citet{Tingay2013OnFeasibility} predicts that debris of radius > $0.5$\,m should be detected with the MWA Phase 1 configuration up to a range of $1000$\,km, for a $50$\,kHz bandwidth and one second integration. Hence, in order to do a basic check of detection completeness, the TLE catalog was used to identify all the objects in LEO, Middle Earth Orbit (MEO) and Highly Elliptical Orbit (HEO) that passed through MWA's half power beam and had their shortest range during pass to be less than $8000$\,km. The shortest range for these satellites during the pass along with their RCS is plotted in Figure \ref{Fig11} (note that the website used for obtaining RCS values does not mention the frequency used for estimating the RCS and are an order of magnitude guide only.  The RCS at the lower MWA frequencies is likely to be smaller). The region shown in \fcolorbox{black}{customColor1}{\rule{0pt}{1pt}\rule{1pt}{0pt}} are all the satellites with RCS > $0.79$\,m$^2$ (2D projection of a sphere of radius $0.5$\,m), the region shown in \fcolorbox{black}{customColor2}{\rule{0pt}{1pt}\rule{1pt}{0pt}} are all the satellites with shortest range less than $1000$\,km and the predicted detection regime is shown as \fcolorbox{black}{customColor}{\rule{0pt}{1pt}\rule{1pt}{0pt}} (as per \citet{Tingay2013OnFeasibility}).

During the 19.34 minutes of observation a total of 49 unique objects with shortest range less than $1000$\,km passed through the half power beam with 3-6 objects being present in the field of view at any given instant.  From Figure \ref{Fig11}, it can be noted that that there are six objects that satisfy the detection criteria. These objects were identified to be ALOS,  Iridium-65 , and four different upper stage rocket debris (the RCS of the six objects are given in Table \ref{tab2}). But of the six objects only ALOS (the biggest) was detected in the full bandwidth difference images (note that ISS was detected outside the half power beam due to its very large RCS and hence is not part of the objects in Figure \ref{Fig11}). Three of the four rocket bodies are in the observation containing the transmitting satellites and were within the field of view when these CubeSats were found to be transmitting. Hence, it is  possible that these objects went undetected due to the increased noise caused by the side-lobes of the bright transmitting CubeSats (for example, the RMS of the $30.72$\, MHz difference images increased from $0.8$\,Jy to about $6.4$\,Jy when the CubeSats were visible). Other missed detections could be also due to unfavourable reflection geometries or weak reflections confining the signal to very few frequency channels, thus reducing the signal to noise ratio in the $30.72$\,MHz bandwidth images. Alouette-2 on the other hand was detected outside the predicted detection parameter space, possibly due to the existence of two dipole antennas of lengths $22.8$\,m and $73$\,m on the satellite (thus increasing its RCS in radio frequencies) for ionosphere sounding purposes\footnote{\url{https://nssdc.gsfc.nasa.gov/nmc/spacecraft/display.action?id=1965-098A}}.

The \citet{Tingay2013OnFeasibility} simulations assumed a 50 kHz bandwidth and a 1 second integration, rather than the full bandwidth 30.72 MHz noted above, which combines a lot of frequency channels with no signal and dilutes the narrow band signal, reducing sensitivity.  Also, the simulations assume detection in the maximum sensitivity pointing direction, whereas all the objects in Figure \ref{Fig11} lie away from this direction to various degrees.  Thus, all of these effects plausibly explain why we detect only a subset of these objects. The minimum angular distance from the pointing center for the 5 undetected objects and the detected satellites is given in Table \ref{tab2}.

Upon detailed inspection of difference images made for fine channels that overlap with known FM frequencies, a transient signal (such as the one shown in Figure \ref{Fig9UNCATALOG}) was found near the predicted location of Iridium-65 and was found to be reflecting in FM frequencies using the DSNRS analysis. But due to the event being confined to a single time-step, the signal could not be definitively identified to be Iridium-65 by comparing its trajectory with the predicted trajectory of the satellite.

\begin{table}
    \caption{The table gives the Boresight angle (denoted as $\theta$) of the 5 undetected objects along with the detected satellites (marked with asterisk) from the pointing center. It also gives the minimum range to target and the RCS for each of the considered objects.}
    \centering
    \begin{tabular}{@{}cccc@{}}
    \hline \hline
    Object  & $\theta$  & Range  &RCS  \\
         &  (Degrees) & (km) & (m$^2$) \\
          \hline \hline
        IRIDIUM-65 & 11.4 & 802 &3.6\\
        \hline
        DELTA 1 R/B & 12.4 & 783 &8.9\\
        \hline
        DELTA 2 R/B & 16.2 & 701 &9.5\\
        \hline
        SL-8 R/B(1) & 16.1 & 735 &4.2\\
        \hline
        SL-8 R/B(2)  & 18.4 &  811 &4.4\\
        \hline
        Alouette-2$^{\ast}$   & 12.5  & 2191 &1.0\\
        \hline
        ALOS$^{\ast}$ & 9.7 & 715&13.6\\
        \hline
        UKube-1$^{\ast}$ & 14.4 & 644&0.1\\
        \hline
        ISS$^{\ast}$  & 26.2& 437&399.2\\
        \hline
        Duchifat-1$^{\ast}$ & 12.6 & 624 & 0.03\\
        \hline \hline
    \end{tabular}
    \label{tab2}
\end{table}

\subsection{Future work}
The DSNRS technique developed here helps classify signals based on their reflection/transmission spectra and the results we obtain support the idea of using the MWA for space surveillance due to its wide-field view. The TLE time offset mentioned in Table \ref{tab1} can be used to update the satellite catalog in the future. Given that many satellites transmit at about $145$\,MHz, observing in these frequencies can help expand our detection window to also include transmitting CubeSats as well. In the future, blind detection of satellites can also be done with higher sensitivity using the compact configuration of Phase 2 of the MWA \citep{2018PASA...35...33W}. The compact configuration has two dense cores with most of the baselines being shorter than $200$\,m, thus being ideal for performing near-field detections. 

Future observations and data processing are planned in order to systematically assess the sensitivity of the techniques developed in this paper.  For example, the MWA Voltage Capture System (VCS) \citep{2015PASA...32....5T} was recently used to perform data collection during the so-called SpaceFest2 event \footnote{\url{https://www.airforce.gov.au/our-mission/spacefest-edge}}, coordinated by the Australian Department of Defence in order to evaluate different technologies and sensor types for SSA . These observations were designed for coherent passive radar with the MWA, utilising processing as described by Hennessy et al. (2019, in press).  However, we have used an offline correlation system for the MWA to convert the captured voltages into visibility datasets suitable for the non-coherent techniques developed in this paper. Thus, from the SpaceFest2 observations, we will be able to compare and test the limits of both coherent and non-coherent techniques with the MWA, from targeted observations (as opposed to the surreptitious observations used in this paper) over a range of objects with different RCS values.

\section{SUMMARY}

In this paper we have:
\begin{itemize}
    \item undertaken a detailed analysis of the apparent FM reflections from LEO RFI detected in \citet{Zhang2018LimitsMWA};
    \item developed an analysis using a quantity we call the Dynamic Signal to Noise Ratio Spectrum (DSNRS), that helps classify signals as originating from objects in orbit from terrestrial transmitters;
    \item used the DSNRS to analyse three signals found to be FM reflections from obejcts in orbit (up-to a maximum range of $2298$\, km) and signals identified as out-of-band transmissions from two CubeSats;
    \item detected short duration signals at FM frequencies that do not coincide with satellite locations predicted using the TLE catalog;
    \item identified position off-set between the actual and predicted TLE position, thus demonstrating MWA's potential to be used for catalog maintenance;
    \item performed a study completeness analysis, considering the reasons why some satellites in the field-of-view went undetected during the observations used in this work.
\end{itemize}

\begin{acknowledgements}

This scientific work makes use of the Murchison Radio-astronomy Observatory, operated by CSIRO. We acknowledge the Wajarri Yamatji people as the traditional owners of the Observatory site. Support for the operation of the MWA is provided by the Australian Government (NCRIS), under a contract to Curtin University administered by Astronomy Australia Limited. We acknowledge the Pawsey Supercomputing Centre which is supported by the Western Australian and Australian Governments.

\subsection{Sofware}
We acknowledge the work and the support of the developers of the following Python packages:
Astropy \citep{theastropycollaboration_astropy_2013,astropycollaboration_astropy_2018}, Numpy \citep{vanderwalt_numpy_2011}, Scipy  \citep{jones_scipy_2001}, matplotlib \citep{Hunter:2007} and Ephem\footnote{\url{https://pypi.org/project/ephem/}}. The work also used WSCLEAN \citep{offringa-wsclean-2014,offringa-wsclean-2017} for making fits images and DS9\footnote{\href{http://ds9.si.edu/site/Home.html}{ds9.si.edu/site/Home.html}} for visualization purposes. BANE \citep{Hancock_Aegean_2012,Hancock_Aegean2_2018} was used for calculating the integrated flux densities and for performing noise estimations.

%The python script used in this work uses the Astropy\footnote{\url{https://www.astropy.org/}}, Ephem\footnote{\url{https://pypi.org/project/ephem/}}, matplotlib\footnote{\url{https://matplotlib.org/}} and numpy\footnote{\url{https://www.numpy.org/}}  modules. WSCLEAN \citep{offringa-wsclean-2014,offringa-wsclean-2017} was used to make the fits images and DS9\footnote{\url{http://ds9.si.edu/site/Home.html}} was used to view them. Aegean \citep{Hancock_Aegean_2012,Hancock_Aegean2_2018} was used for calculating the integrated flux densities and for performing noise estimation.

\end{acknowledgements}

%\end{appendix}

\bibliographystyle{pasa-mnras}
\bibliography{custom}

\end{document}